\documentclass[12pt]{article}

\usepackage{latexsym}
\usepackage{amssymb}
\usepackage{amsmath}
\def\i{{\rm i}}
\def\e{{\rm e}}
\def\d{{\rm d}}
\def\one{{\sf 1}\mkern-5.0mu{\rm I}}
\title{Asymptotic Form of Zero Energy Wave Functions in 
Supersymmetric Matrix Models}
\author{J. Fr\"ohlich${}^{(a)}$ , G.M. Graf${}^{(a)}$ , D. Hasler${}^{(a)}$ , 
J. Hoppe${}^{(b,c)}$ , S.-T. Yau${}^{(d)}$ \\ 
\vspace*{-0.05truein} \\
\normalsize\it ${}^{(a)}$ Theoretische Physik,
ETH-H\"onggerberg, CH--8093 Z\"urich\\
\normalsize\it \hskip -0.63cm ${}^{(b)}$ Max-Planck-Institut f\"ur 
Gravitationsphysik,
Albert-Einstein-Institut, D-14473 Potsdam \\
\normalsize\it ${}^{(c)}$ Fachbereich Mathematik, TU Berlin, D-10623 Berlin\\
\normalsize\it ${}^{(d)}$ Department of Mathematics, Harvard University, 
Cambridge, MA 02138}

\begin{document}

\maketitle
\vspace{0.4cm}
\begin{abstract}
We derive the power law decay, and asymptotic form, of 
${\rm SU}(2)\times{\rm Spin}(d)$ invariant wave--functions satisfying
$Q_\beta\psi=0$ for all $s_d=2(d-1)$ supercharges of reduced
$(d+1)$--dimensional supersymmetric ${\rm SU}(2)$ Yang Mills theory, resp. of 
the ${\rm SU}(2)$--matrix model related to supermembranes in $d+2$ dimensions.
\end{abstract}

\section{Introduction}
\label{sec:intr}

It is generally believed that supersymmetric ${\rm SU}(N)$ matrix models 
in $d=9$
dimensions admit exactly one normalizable zero-energy solution for each $N>1$,
while they admit none for all other dimensions for which the models can be
formulated, i.e., for $d=2,3,5$. For various approaches to this problem see 
e.g. \cite{c1a}--\cite{c1l}. 

In this article, we would like to summarize (and slightly modify/extend) what
is known about the behaviour of ${\rm SU}(2)$ zero--energy solutions far out at
infinity in (and near) the space of configurations where the bosonic potential
(the trace of all commutator--squares) vanishes. Based on some early 'negative'
result concerning $N=2,\, d=2$ (that used rather different 
techniques/arguments; see \cite{c1a, c4a})
we started our investigation of the asymptotic behaviour, in the fall of 1997,
with a Hamiltonian Born--Oppenheimer analysis of that $N=2,\, d=2$ case. 
Some months later,
we realized that the rather complicated Hamiltonian analysis (Halpern
and Schwartz \cite{c1h} had, in the meantime, derived the form of the wave 
function for $d=9$ near $\infty$, by Hamiltonian Born--Oppenheimer methods) 
can be replaced by a
simple first order analysis, using only the first order operators $Q$,
and first order perturbation theory. One finds that asymptotically
normalizable, ${\rm SU}(2)$ and ${\rm SO}(d)$ invariant, wave functions do 
not exist for
$d=2,3$, and $5$, in contrast to $d=9$, where there is exactly one.

We close these introductory words by recalling that the models discussed below
arise in at least 3 somewhat different ways: As supersymmetric extensions of
regulated membrane theories in $d+2$ space--time dimensions 
\cite{c2a, c4a},
as reductions (to $0+1$ dimension) of $d+1$ dimensional Super Yang Mills 
theories \cite{c3a}--\cite{c3c}, 
and, for $d=9$, as a description of the dynamics of D--0 branes in superstring
theory, \cite{c5a, c5b}. In this physical interpretation, the existence of a 
normalizable zero--energy solution is an important consistency requirement. 

The paper is organized as follows. In Section \ref{sec:mod} we recall the 
definition of the models, and in Section \ref{sec:res} we state our main 
result about zero--modes. The proof is given in Section \ref{sec:pr} and
Appendix 1. We suggest to skip Subsection \ref{sec:pr5} and Appendix 1 at a 
first reading. As a warm--up the
reader is advised to read Appendix 2, where a simpler model is
treated by the same method.

\section{The models}
\label{sec:mod}

The configuration space of the bosonic degrees of freedom is $X=\mathbb R^{3d}$
with coordinates 
\begin{equation}
q=(\vec{q_1},\ldots,\vec{q_d})=(q_{sA})_{\substack{s=1,\ldots, d\\A=1,2,3}}\;.
\nonumber
\end{equation}
To describe the fermionic degrees of freedom let, as a preliminary,
\begin{equation}
\gamma^i=(\gamma^i_{\alpha\beta})_{\alpha,\beta=1,\ldots,s_d}\;,\qquad
(i=1,\ldots,d)\;,
\label{e1.1aa}
\end{equation}
be the {\it real} representation of smallest dimension, called $s_d$, of the 
Clifford algebra with $d$ generators: 
$\{\gamma^s,\gamma^t\}=2\delta^{st}\one$. On the representation space, 
${\rm Spin}(d)$ is realized through matrices 
$R\in{\rm SO}(s_d)$, so that we may view 
\begin{equation} 
{\rm Spin}(d)\hookrightarrow{\rm SO}(s_d)\;,
\label{e1.1a}
\end{equation}
as a simply connected subgroup. We recall that
\begin{equation*}
s_d=\cases 2^{[d/2]}\;,& d=0,1,2 \mod 8\;,\\
          2^{[d/2]+1}\;,& \hbox{otherwise}\;,
    \endcases
\end{equation*}
where $[\cdot]$ denotes the integer part. We then consider the Clifford 
algebra with $s_d$ generators and its irreducible representation on 
$\mathcal C=\mathbb C^{2^{s_d/2}}$. On 
$\mathcal C^{\otimes 3}$ the Clifford generators 
\begin{equation}
(\vec{\Theta}_1,\ldots,\vec{\Theta}_{s_d})=
(\Theta_{\alpha A})_{\substack{\alpha=1,\ldots,s_d\\A=1,2,3}}
\nonumber
\end{equation}
are defined, satisfying 
$\{\Theta_{\alpha A} ,\Theta_{\beta B}\}=\delta_{\alpha\beta}\,\delta_{AB}$.
The Hilbert space, finally, is 
\begin{equation}
\mathcal H={\rm L}^2(X,\mathcal C^{\otimes 3})\;.
\label{e1.0}
\end{equation}

There is a natural representation of ${\rm SU}(2)\times{\rm Spin}(d)\ni(U,R)$ 
on $\mathcal H$. In fact, the group acts naturally on $X$ through its
representation ${\rm SO}(3)\times{\rm SO}(d)$ (which we also denote by
$(U,R)$). On $\mathcal C^{\otimes 3}$
we have the representation $\mathcal R$ of ${\rm Spin}(s_d)\ni R$
\begin{equation} 
\mathcal R(R)^*\Theta_{\alpha A}\mathcal R(R)=\widetilde{R}_{\alpha\beta}
\Theta_{\beta A}\;,
\label{e1.2}
\end{equation}
where $\widetilde{R}=\widetilde{R}(R)$ is its ${\rm SO}(s_d)$ representation.
Through ${\rm SO}(s_d)={\rm Spin}(s_d)/\mathbb Z_2$ and (\ref{e1.1a}) we have
\begin{equation} 
{\rm Spin}(d)\hookrightarrow{\rm Spin}(s_d)\;,
\label{e1.2a}
\end{equation}
and thus a representation $\mathcal R$ of ${\rm Spin}(d)$. The representation 
$\mathcal U$ of ${\rm SU}(2)\ni U$ on $\mathcal C^{\otimes 3}$ is
characterized by
$\mathcal U(U)^*\Theta_{\alpha A}\mathcal U(U)=U_{AB}\Theta_{\alpha B}$.

We shall now restrict to $d=2,\,3,\,5,\,9$, where $s_d=2,\,4,\,8,\,16$, the 
reason being that in these cases  
\begin{equation}
s_d=2(d-1)\;,
\label{e1.1}
\end{equation}
whereas $s_d$ is strictly larger otherwise. Eq. (\ref{e1.1}) is essential
for the algebra (\ref{eq2}) below \cite{c3c}.

The supercharges, acting on $\mathcal H$, are given by the $s_d$ hermitian 
operators
\begin{equation} 
Q_\beta= \vec{\Theta}_\alpha \cdot \bigl( 
-\i \gamma^t_{\alpha\beta} \vec{\nabla}_t + \frac{1}{2} \,  
\vec{q}_s \times \vec{q}_t\,\gamma^{st}_{\beta\alpha} \bigr)\;, 
\qquad(\beta=1,\ldots, s_d)\;,
\nonumber
\end{equation}
where $\gamma^{st}=(1/2)(\gamma^s \gamma^t -\gamma^t \gamma^s)$. These
supercharges transform as scalars under 
${\rm SU}(2)$ transformations generated by 
\begin{equation} 
J_{AB}=-\i(q_{sA}\partial_{sB}-q_{sB}\partial_{sA})-\frac{\i}{2}
(\Theta_{\alpha A}\Theta_{\alpha B}-\Theta_{\alpha B}\Theta_{\alpha A})
\equiv L_{AB}+M_{AB}\;,
\nonumber
\end{equation}
resp. as vectors in $\mathbb R^{s_d}$ under ${\rm Spin}(d)$ transformation 
generated by 
\begin{equation} 
J_{st} = -\i(\vec{q}_s\cdot\vec{\nabla}_t - 
\vec{q}_t\cdot\vec{\nabla}_s)-\frac{\i}{4} 
\,\vec{\Theta}_\alpha \gamma^{st}_{\alpha\beta} \vec{\Theta}_\beta
\equiv L_{st}+M_{st}\;. 
\nonumber
\end{equation}
The anticommutation relations of the supercharges are
\begin{equation} 
\bigl\{Q_\alpha, Q_\beta\bigr\}=\delta_{\alpha\beta}H+
\gamma^t_{\alpha\beta}q_{tA}\varepsilon_{ABC}J_{BC}\;.
\label{eq2}
\end{equation}
Here, $H$ is the Hamiltonian
\begin{equation} 
H = - \sum_{s=1}^9 \, \vec{\nabla}_s^2 + \sum_{s<t} 
\bigl( \vec{q}_s \times  \vec{q}_t \bigr)^2 + 
\i \vec{q}_s \cdot\bigl(\vec{\Theta}_\alpha \times \vec{\Theta}_\beta\bigr) 
\, \gamma^s_{\alpha\beta}  \; ,
\label{e1.4} 
\end{equation}
which commutes with both $J_{AB}$ and $J_{st}$. The question we address is the
possibility of a normalizable state $\psi\in\mathcal H$ with zero energy,
i.e., with $H\psi=0$, which is a singlet w.r.t. both ${\rm SU}(2)$ and 
${\rm Spin}(d)$. Note that on  ${\rm SU}(2)$ invariant states 
$H=2Q_\beta^2\ge 0$ and in fact the energy spectrum is 
(\cite{c4b}) $\sigma(H)=[0,\infty)$. Equivalently, we look for zero-modes 
\begin{equation} 
Q_\beta\psi=0\;,\qquad(\beta=1,\ldots, s_d)\;.
\nonumber
\end{equation}

\section{Results}
\label{sec:res}

The potential $\sum_{s<t}( \vec{q}_s \times  \vec{q}_t)^2$ vanishes on the 
manifold 
\begin{equation} 
\vec{q}_s = r \vec{e} E_s 
\nonumber
\end{equation}
with $r>0$ and $\vec{e}{\,}^2=\sum_s E_s^2 = 1$. The dimension of the manifold is 
$1+2+(d-1)=3d-2(d-1)$. Points in a conical neighborhood
of the manifold can be expressed in terms of tubular (or ``end--point'') 
coordinates \cite{b2}
\begin{equation} 
\vec{q}_s = r \vec{e} E_s + r^{-1/2}\vec{y}_s 
\label{e2.00}
\end{equation}
with 
\begin{equation} 
\vec{y}_s\cdot \vec{e} = 0\;,\qquad\vec{y}_s E_s =\vec{0}\;.
\label{e2.1} 
\end{equation}
A prefactor 
has been put explicitely in front of the transversal coordinates $\vec{y}_s$, 
so as to anticipate the length scale $r^{-1/2}$ of the ground state. 
The change 
\begin{equation} 
(\vec{e},E, y)\mapsto(-\vec{e},-E, y)
\label{e2.0}
\end{equation}
does not affect $\vec{q}_s$. Rather than identifying the two coordinates for
$\vec{q}_s$,
we shall look for states which are even under the antipode map 
(\ref{e2.0}).

\smallskip
We can now describe the structure of a putative ground state.

\smallskip\noindent
{\bf Theorem} {\it Consider the equations $Q_\beta\psi=0$ for a formal power 
series solution near $r=\infty$ of the form
\begin{equation} 
\psi=r^{-\kappa}\sum_{k=0}^\infty r^{-\frac{3}{2}k}\psi_k\;,
\label{e2.4} 
\end{equation}
where: $\psi_k=\psi_k(\vec{e},E,y)$ is square integrable w.r.t. 
$\d e\,\d E\,\d y$;
\newline
\hphantom{where:} $\psi_k$ is ${\rm SU}(2)\times{\rm Spin}(d)$ 
invariant;\newline
\hphantom{where:} $\psi_0\neq 0$.\newline
Then, up to linear combinations,
\begin{itemize}
\item{d=9}: The solution is unique, and $\kappa=6$;
\item{d=5}: There are three solutions with $\kappa=-1$ and one 
with $\kappa=3$;
\item{d=3}: There are two solutions with $\kappa=0$;
\item{d=2}: There are no solutions.
\end{itemize}
All solutions are even under the antipode map (\ref{e2.0}),
\begin{equation*} 
\psi_k(\vec{e},E,y)=\psi_k(-\vec{e},-E,y)\;,
\end{equation*}
except for the state $d=5,\,\kappa=3$, which is odd.}

\medskip\noindent
{\bf Remarks} 1. The equation $Q_\beta\psi=0$ can be viewed as an ordinary
differential equation in $z=r^{3/2}$ for a function taking values in 
${\rm L}^2(\d e\,\d E\,\d y,\mathcal C^{\otimes 3})$ (see eq. (\ref{e3.00}) 
below). It turns out that
$z=\infty$ is a singular point of the second kind \cite{b1}. In such a 
situation the series (\ref{e2.4}) is typically asymptotic to a true solution,
but not convergent.
\par\noindent
2. The integration measure is 
$\d q=\d r\cdot r^2\d e\cdot r^{d-1}\d E\cdot r^{-\frac{1}{2}\cdot 2(d-1)}\d y=
r^2\d r\,\d e\,\d E\,\d y$. The wave function (\ref{e2.4}) is square 
integrable 
at infinity if 
$\int^\infty\d r\,r^2(r^{-\kappa})^2<\infty$, i.e., if $\kappa>3/2$. The 
theorem is consistent with the statement according to which {\bf only} for 
$d=9$ a (unique) normalizable ground state for (\ref{e1.4}) (which would
have to be even) is possible.
\par\noindent
3. Note that the connection of matrix models with supergravity requires the
zero--energy solutions to be ${\rm Spin}(d)$ singlets only for $d=9$.

\bigskip
The case $d=2$ can be dealt with immediately. We may assume 
$\gamma^2 =\sigma_3, \,
\gamma^1 = \sigma_1$ (Pauli matrices), so that
\begin{equation*}
M_{12}=\frac{\i}{2}\Theta_{1A}\Theta_{2A}\;,
\end{equation*}
with commuting terms. Since, for each $A=1,2,3,\, (\Theta_{1A}\Theta_{2A})^2
=-1/4$, we see that $M_{12}$ has spectrum in $\mathbb Z/2 +1/4$. Given that 
$L_{12}$ has spectrum $\mathbb Z$, no state with $J_{12}\psi=0$ is possible.
We mention \cite{c1a} that, more generally, for $d=2$ no
normalizable ${\rm SU}(2)$ invariant ground state exists.

The proof of the theorem will thus deal with $d=9,5,3$ only.

\section{Proof}
\label{sec:pr}

We shall first derive the power series expansion of the supercharges
$Q_\beta$. To this end we note that
\begin{eqnarray}
\frac{\partial}{\partial q_{tA}}&=&
r^{1/2}(\delta_{st}-E_sE_t)(\delta_{AB}-e_Ae_B)
\frac{\partial}{\partial y_{sB}}
\label{e3.000}\\
&&+r^{-1}[e_AE_t(r\frac{\partial}{\partial r}+
\frac{1}{2}y_{sB}\frac{\partial}{\partial y_{sB}})+
\i e_BE_tL_{BA}+\i e_AE_sL_{st}]
+{\rm O}(r^{-5/2})\;,\nonumber
\end{eqnarray}
with the remainder not containing derivatives w.r.t. $r$ (see Appendix 1 for 
derivation). This yields
\begin{equation} 
Q_\beta=r^{1/2}Q_\beta^0+
r^{-1}(\widehat{Q}_\beta^1r\frac{\partial}{\partial r}+Q_\beta^1)
+r^{-5/2}Q_\beta^2+\ldots
\label{e3.00}
\end{equation}
with $r$--independent operators
\begin{eqnarray*}
Q_\beta^0&=&-\i\Theta_{\alpha A}\gamma^t_{\alpha\beta}(\delta_{st}-E_sE_t)
(\delta_{AB}-e_Ae_B)\frac{\partial}{\partial y_{sB}}+
\vec{\Theta}_\alpha
\cdot(\vec{e}\times\vec{y}_t)E_s\gamma^{st}_{\beta\alpha}\;,\\
\widehat{Q}_\beta^1&=&
-\i (\vec{\Theta}_\alpha\cdot\vec{e}\,)\gamma^t_{\alpha\beta}E_t\;,\\
Q_\beta^1&=&\Theta_{\alpha A}\gamma^t_{\alpha\beta}
\bigl(e_BE_tL_{BA}+
e_AE_sL_{st}-\frac{\i}{2}\,e_AE_ty_{sB}\frac{\partial}{\partial y_{sB}})+
\frac{1}{2}\vec{\Theta}_\alpha\cdot(\vec{y}_s\times\vec{y}_t)
\gamma^{st}_{\beta\alpha}\;.
\end{eqnarray*}
The explicit expressions of $Q_\beta^n,\, (n\ge 2)$ will not be needed. We
then equate coefficients of powers of $r^{-3/2}$ in the equation 
$Q_\beta\psi=0$ with the result
\begin{eqnarray}
Q_\beta^0\psi_n+
\bigl(-(\kappa+\frac{3}{2}(n-1))\widehat{Q}_\beta^1+Q_\beta^1\bigr)
\psi_{n-1}+Q_\beta^2\psi_{n-2}+\ldots+Q_\beta^n\psi_0&=&0\;,\nonumber\\
\qquad(n&=&0,1,\ldots)\;.
\label{e3.0}
\end{eqnarray}

\subsection{The equation at $n=0$}
\label{sec:pr1}

The equation at $n=0$,
\begin{equation} 
Q_\beta^0\psi_0=0\;,
\label{e3.1}
\end{equation}
admits precisely the (not necessarily ${\rm SU}(2)\times{\rm Spin}(d)$
invariant) solutions 
\begin{equation} 
\psi_0(\vec{e},E, y)=\e^{-\sum_s\vec{y}_s{\,}^2/2}|F(E,\vec{e})\rangle\;,
\label{e3.2}
\end{equation}
(with $\vec{y}$ restricted to (\ref{e2.1})), where the fermionic states 
$|F(E,\vec{e})\rangle$ can be described as follows:
Let $\vec{n}_\pm$ be two complex vectors satisfying 
$\vec{n}_+\cdot\vec{n}_-=1,\,\vec{e}\times \vec{n}_\pm=\mp\i\vec{n}_\pm$ (and 
hence 
$\vec{n}_\pm\cdot\vec{n}_\pm=0$, $\vec{n}_+\times\vec{n}_-=-\i\vec{e}$\,). For 
any vector $v\in\mathbb R^{s_d}$ 
we may introduce $\vec{\Theta}(v)=\vec{\Theta}_\alpha v_\alpha$, as well as
fermionic operators $\vec{\Theta}(v)\cdot\vec{n}_\pm$ satisfying
canonical anticommutation relations:
\begin{equation} 
\bigl\{\vec{\Theta}(u)\cdot\vec{n}_+,\vec{\Theta}(v)\cdot\vec{n}_-\bigr\} 
=u_\alpha v_\alpha\;,\qquad
\bigl\{\vec{\Theta}(u)\cdot\vec{n}_\pm,\vec{\Theta}(v)\cdot\vec{n}_\pm\bigr\} 
=0\;.
\nonumber
\end{equation}
Then, $|F(E,\vec{e})\rangle$ is required to obey
\begin{equation} 
\vec{\Theta}(v)\cdot\vec{n}_\pm|F(E,\vec{e})\rangle=0
\qquad\hbox{for}\qquad
E_s\gamma^sv=\pm v\;.
\label{e3.4}
\end{equation}
To prove the above, let us note that 
\begin{eqnarray}      
&&\hskip 1cm\bigl\{Q_\alpha^0,Q_\beta^0\bigr\}=
\delta_{\alpha\beta}H^0+
\gamma^t_{\alpha\beta}E_t\varepsilon_{ABC}M_{AB}e_C\;,
\label{e3.4a}\\
 H^0&=&\bigl[-(\delta_{st}-E_sE_t)(\delta_{AB}-e_Ae_B)
\frac{\partial}{\partial y_{sA}}\frac{\partial}{\partial y_{tB}}
+\sum_s\vec{y}_s^2\bigr]+
\i E_s\gamma^s_{\alpha\beta}\vec{e}\cdot
\bigl(\vec{\Theta}_\alpha \times \vec{\Theta}_\beta\bigr)\nonumber\\
&\equiv& H^0_B+H^0_F\;.\nonumber
\end{eqnarray}
By contracting eq. (\ref{e3.4a}) against $\delta_{\alpha\beta}$, resp. 
$\gamma^t_{\alpha\beta}E_t$ we see that the equations (\ref{e3.1}) are 
equivalent to the pair of equations
\begin{equation} 
H^0\psi_0=0\;,\qquad \varepsilon_{ABC}M_{AB}e_C\psi_0=0\;.
\label{e3.4b}
\end{equation}
Here, $H^0_B$ is a harmonic oscillator in $2(d-1)$ degrees of
freedom, with orbital ground state wave function 
$\e^{-\sum_s\vec{y}_s^2/2}$ and energy $2(d-1)$. On the other hand,
\begin{eqnarray}
H^0_F&=&-E_s\gamma^s_{\alpha\beta}
\bigl((\vec{\Theta}_\alpha\cdot\vec{n}_+)(\vec{\Theta}_\beta\cdot\vec{n}_-)-
(\vec{\Theta}_\alpha\cdot\vec{n}_-)(\vec{\Theta}_\beta\cdot\vec{n}_+)\bigr)
\nonumber\\
&=&-s_d+
2P^+_{\alpha\beta}(\vec{\Theta}_\alpha\cdot\vec{n}_-)
(\vec{\Theta}_\beta\cdot\vec{n}_+)
+2P^-_{\alpha\beta}(\vec{\Theta}_\alpha\cdot\vec{n}_+)
(\vec{\Theta}_\beta\cdot\vec{n}_-)\;,
\label{e3.5}
\end{eqnarray}
where we used the spectral decomposition $E_s\gamma^s=P^+-P^-$. In view of 
(\ref{e1.1}), the equation $H^0\psi_0=0$ is fulfilled iff the fermionic
state is annihilated by the last two positive terms in (\ref{e3.5}), i.e., if 
(\ref{e3.4}) holds. The second equation (\ref{e3.4b}) is now also satisfied,
since 
\begin{eqnarray}
\frac{1}{2}\varepsilon_{ABC}M_{AB}e_C&=&
-\frac{\i}{2}\vec{e}\cdot
\bigl(\vec{\Theta}_\alpha \times \vec{\Theta}_\alpha \bigr)\nonumber\\
&=&\frac{1}{2}
\bigl((\vec{\Theta}_\alpha\cdot\vec{n}_+)(\vec{\Theta}_\alpha\cdot\vec{n}_-)-
(\vec{\Theta}_\alpha\cdot\vec{n}_-)(\vec{\Theta}_\alpha\cdot\vec{n}_+)\bigr)
\nonumber\\
&=&P^-_{\alpha\beta}(\vec{\Theta}_\alpha\cdot\vec{n}_+)
(\vec{\Theta}_\beta\cdot\vec{n}_-)-
P^+_{\alpha\beta}(\vec{\Theta}_\alpha\cdot\vec{n}_-)
(\vec{\Theta}_\beta\cdot\vec{n}_+)
\label{e3.5c}
\end{eqnarray}
annihilates $|F(E,\vec{e})\rangle$.
\par

\subsection{${\rm SU}(2)\times{\rm Spin}(d)$ invariant states}
\label{sec:pr2}

We recall that the representation $\mathcal R[\cdot]$ of ${\rm Spin}(d)$ on 
$\mathcal H$ is 
$(\mathcal R[R]\psi)(q)=\mathcal R(R)(\psi(R^{-1}q))$, where $\mathcal R(R)$
acts on $\mathcal C^{\otimes 3}$. Similarly for ${\rm SU}(2)$.
The invariant solutions among (\ref{e3.2}) are thus 
those which satisfy
\begin{equation} 
\mathcal U(U)|F(E,\vec{e})\rangle =|F(E,U\vec{e})\rangle\;,\qquad
\mathcal R(R)|F(E,\vec{e})\rangle =|F(RE,\vec{e})\rangle\;,
\label{e3.5a}
\end{equation}
for $(U,R)\in{\rm SU}(2)\times{\rm Spin}(d)$. These states are in bijective 
correspondence to states invariant under the `little group' 
$(U,R)\in{\rm U}(1)\times{\rm Spin}(d-1)$, i.e., to states 
$|F(E,\vec{e})\rangle$ satisfying
\begin{equation} 
\mathcal U(U)|F(E,\vec{e})\rangle =|F(E,\vec{e})\rangle\;,\qquad
\mathcal R(R)|F(E,\vec{e})\rangle =|F(E,\vec{e})\rangle\;,
\label{e3.5b}
\end{equation}
for some arbitrary but fixed $(E,\vec{e})$ and all $U,\,R$ with 
$U\vec{e}=\vec{e},\,RE=E$. The first relation holds on all of (\ref{e3.4}). In
fact the generator (\ref{e3.5c}) of the group $\mathcal U(U)$ of rotations 
$U$ about $\vec{e}$ annihilates $|F(E,\vec{e})\rangle$, as we just saw. To 
discuss the second relation 
(\ref{e3.5b}) we note that the generators of ${\rm Spin}(d-1)$ (i.e., of the 
fermionic rotations about $E$), are $M_{st}U_sV_t$ with $U_sE_s=V_sE_s=0$. We 
write
$M_{st}=M_{st}^{\perp}+M_{st}^{\parallel}$, 
where 
\begin{equation} 
M_{st}^{\perp}=-(\i/2)(\vec{\Theta}_\alpha\cdot\vec{n}_+)
\gamma^{st}_{\alpha\beta}(\vec{\Theta}_\beta\cdot\vec{n}_-)\;,\qquad
M_{st}^{\parallel}=-(\i/4)(\vec{\Theta}_\alpha\cdot\vec{e}\,)
\gamma^{st}_{\alpha\beta}(\vec{\Theta}_\beta\cdot\vec{e}\,)\;,
\label{e3.8b}
\end{equation}
and remark that, by a computation similar to (\ref{e3.5c}), 
$M_{st}^{\perp}U_sV_t$ annihilates $|F(E,\vec{e})\rangle$. As a result, we may
study the representation $\mathcal R$ of the group ${\rm Spin}(d-1)$ through 
its embedding in the Clifford algebra generated by the 
$\vec{\Theta}_\alpha\cdot\vec{e}$.
\par
The operators $\vec{\Theta}_\alpha\cdot\vec{e}$ leave the space (\ref{e3.4})
invariant and act irreducibly on it. That space is thus isomorphic to 
$\mathcal C$, and ${\rm Spin}(s_d)$ acts according to (\ref{e1.2}) (with 
$\Theta_{\alpha A}$ replaced by $\vec{\Theta}_\alpha\cdot\vec{e}$). This 
representation decomposes (see e.g. \cite{b3}) as 
\begin{equation}
\mathcal C=(2^{(s_d/2)-1})_+\oplus (2^{(s_d/2)-1})_-
\label{e3.2a}
\end{equation}
w.r.t. the subspaces where 
$\Theta\equiv 
2^{s_d/2}\prod_{\alpha=1}^{s_d}\vec{\Theta}_\alpha\cdot\vec{e}=+1$,
resp. $-1$. The embedding (\ref{e1.2a}) and the corresponding branching of the
representation (but not the statement of the theorem!) depend on the choice
of the $\gamma$--matrices. In order to select a definite embedding, let
\begin{equation} 
\gamma^d =\left( 
\begin{array}{cc}  \one & \ 0 \\ 0 & - \one \end{array} \right) \; , \quad 
\gamma^{d-1} = \left( 
\begin{array}{cc}  \ 0 &\ \one  \\ \ \one & \ 0\end{array} \right) \; , \quad  
\gamma^j =\left( 
\begin{array}{cc}  0 & \i \Gamma^j\\  -\i \Gamma^j& 0 \end{array} \right) \;
\label{e3.3}
\end{equation}
with $\Gamma^j,\,(j= 1,\ldots, d-2)$ purely imaginary, antisymmetric, and
$\{ \Gamma^j, \Gamma^k \} = 2 \delta_{jk} \one_{s_d/2}$. Then
(\ref{e3.2a}) branches as (see \cite{b4}, resp. \cite{c1k, c1l})
\begin{equation} 
\mathcal C=\cases (44\oplus 84)\oplus 128\;,&\qquad (d=9)\;,\\
     (5\oplus 1\oplus 1\oplus 1)\oplus (4 \oplus 4)\;,&\qquad (d=5)\;,\\
        2\oplus (1\oplus 1) \;,&\qquad (d=3)\;,
           \endcases
\label{e3.6}
\end{equation}
when viewed as a representation of ${\rm Spin}(d)$.  
(The choice 
$\widetilde{\gamma}^i_{\alpha\beta}=
\widetilde{R}_{\alpha'\alpha}\gamma^i_{\alpha'\beta'}
\widetilde{R}_{\beta'\beta}$ with 
$\widetilde{R}\in {\rm O}(s_d),\,\det\widetilde{R}=-1$ would have inverted the
branching of the representations on the r.h.s. of (\ref{e3.2a})). The
case $d=3$ deserves a remark, as there are additional inequivalent embeddings 
${\rm Spin}(d=3)\hookrightarrow{\rm Spin}(s_d=4)$, and one has to consider the 
one appropriate to (\ref{e1.2a}). In fact $R\in{\rm Spin}(3)={\rm SU}(2)$ acts 
in the fundamental representation on $\mathbb C^2$, the irreducible
representation space of the complex Clifford algebra with $3$ generators. The
real representation (\ref{e3.3}) is obtained by joining two complex
representations, followed by an appropriate change $T$ of basis. The embedding 
(\ref{e1.2a}) is thus realized through $R \mapsto
T^{-1}(R\otimes\one_2)T$ and the embedding
${\rm su(2)}_{\mathbb C}\hookrightarrow{\rm so(4)}_{\mathbb C}
={\rm su(2)}_{\mathbb C}\oplus{\rm su(2)}_{\mathbb C}$ is equivalent to
$u\mapsto(u,0)$.
\par
The further branching 
${\rm Spin}(d)\hookleftarrow{\rm Spin}(d-1)$ yields
\begin{equation} 
\mathcal C=\cases (1\oplus 8_{\rm v}\oplus 35_{\rm v})
              \oplus(28\oplus 56_{\rm v}) 
      \oplus(8_{\rm s}\oplus 8_{\rm c}\oplus 56_{\rm s}\oplus 56_{\rm c})
 \;,&\qquad (d-1=8)\;,\\
           1\oplus 1\oplus 1\oplus (1\oplus 4)\oplus 
     (2_+\oplus 2_-)\oplus(2_+\oplus 2_-)
 \;,&\qquad (d-1=4)\;,\\
           (1_1\oplus 1_{-1})\oplus 1_0\oplus 1_0 \;,&\qquad (d-1=2)\;.
           \endcases
\label{e3.7}
\end{equation}
The content of invariant states stated in the theorem is now manifest. One 
should notice that for
$d=3$ the little group ${\rm U}(1)$ is abelian and 
the singlets $1_{\pm 1}$ do not correspond to invariant states. For later 
use we also retain the fermionic ${\rm Spin}(d)$ representation to which the 
remaining singlets are associated,
\begin{equation} 
44\quad(d=9)\;;\qquad 1, 1, 1, 5\quad(d=5)\;;\qquad 1,1\quad(d=3)\;,
\label{e3.8}
\end{equation}
together with the corresponding eigenvalue of $\Theta$:
\begin{equation} 
\Theta=\quad 1\quad(d=9)\;;\qquad  1, 1, 1, 1\quad(d=5)\;;
\qquad  -1,-1\quad(d=3)\;.
\label{e3.8a}
\end{equation}

\subsection{Even states}
\label{sec:pr3}

It remains to check which of these states satisfy $|F(-E,-\vec{e})\rangle
=|F(E,\vec{e})\rangle$. Let us begin by noting that by (\ref{e3.5a})
\begin{equation*} 
|F(-E,-\vec{e})\rangle=\e^{\i M_{AB}e_Au_B\pi}\e^{\i M_{st}E_sU_t\pi}
|F(E,\vec{e})\rangle\;,
\end{equation*}
where $\vec{u}\in\mathbb R^3$, resp. $U\in\mathbb R^d$ are unit vectors
orthogonal to $\vec{e}$, resp. $E$. The ${\rm Spin}(d)$ rotation can be
factorized as 
$\e^{\i M_{st}E_sU_t\pi}=
\e^{\i M_{st}^{\perp}E_sU_t\pi}\e^{\i M_{st}^{\parallel}E_sU_t\pi}$.
We claim that 
$\e^{\i M_{st}^{\parallel}E_sU_t\pi}$ $|F(E,\vec{e})\rangle=
\sigma|F(E,\vec{e})\rangle$
with 
\begin{equation} 
\sigma=\quad 1\quad(d=9)\;;\qquad  1, 1, 1, -1\quad(d=5)\;;
\qquad  1,1\quad(d=3)\;.
\label{e3.9}
\end{equation}
The operator represents a rotation 
$R\in{\rm Spin}(d)$ with $RE=-E$ in the representation (\ref{e3.8}). For $d=9$
the latter can be
realized on symmetric traceless tensors $T_{ij},\,(i,j=1,\ldots, 9)$, where 
the ${\rm Spin}(8)$--singlet is $E_iE_j-(1/9)\delta_{ij}$, implying 
$\sigma=1$. For $d=5$, the last representation (\ref{e3.8}) is just the vector
representation, where $\sigma=-1$. As the remaining cases are evident, eq.
(\ref{e3.9}) is proven. A computation using (\ref{e3.3}) and, without loss
$E=(0,\ldots,0,1),\,U=(0,\ldots,1,0)$ shows
\begin{eqnarray*} 
\e^{\i M_{d, d-1}^{\perp}\pi}|F(E,\vec{e})\rangle&=&
\prod_{\alpha=1}^{s_d/2}\e^{[
(\vec{\Theta}_\alpha\cdot\vec{n}_+)
(\vec{\Theta}_{\alpha+s_d/2}\cdot\vec{n}_-)-
(\vec{\Theta}_{\alpha+s_d/2}\cdot\vec{n}_+)
(\vec{\Theta}_\alpha\cdot\vec{n}_-)]\pi/2}|F(E,\vec{e})\rangle\\
&=&\prod_{\alpha=1}^{s_d/2}
(\vec{\Theta}_{\alpha+s_d/2}\cdot\vec{n}_+)
(\vec{\Theta}_\alpha\cdot\vec{n}_-)|F(E,\vec{e})\rangle\equiv
|\overline{F}(E,\vec{e})\rangle\;,\\
\e^{\i M_{AB}e_Au_B\pi}|\overline{F}(E,\vec{e})\rangle&=&
\prod_{\alpha=1}^{s_d}\e^{
(\vec{\Theta}_\alpha\cdot\vec{e})(\vec{\Theta}_\alpha\cdot\vec{u})\pi}
|\overline{F}(E,\vec{e})\rangle\\
&=&(-1)^{s_d/4}\Theta\prod_{\alpha=1}^{s_d/2}
(\vec{\Theta}_\alpha\cdot\vec{n}_+)
(\vec{\Theta}_{\alpha+s_d/2}\cdot\vec{n}_-)|\overline{F}(E,\vec{e})\rangle=
|F(E,\vec{e})\rangle\;,
\end{eqnarray*}
where we used (\ref{e3.8a}) in the last step. Together with (\ref{e3.9}) this 
proves the statement of theorem 
concerning the invariance under (\ref{e2.0}).

\subsection{The equation at $n>0$}
\label{sec:pr4}

We next discuss the equations (\ref{e3.0})${}_n$ with $n\ge 1$. Let $P_0$ be
the orthogonal projection onto the states (\ref{e3.2}), i.e., onto the null
space of $Q_\beta^0$. We replace them with an equivalent pair of
equations, obtained by multiplication of (\ref{e3.0})${}_{n+1}$ with $P_0$, 
resp. of (\ref{e3.0})${}_n$ with $Q_\beta^0$, which
is injective on the range of the complementary projection 
$\overline{P}_0=1-P_0$:
\begin{eqnarray}
P_0\bigl(-(\kappa+\frac{3}{2}n))\widehat{Q}_\beta^1+Q_\beta^1\bigr)P_0\psi_n
=-P_0\bigl(Q_\beta^1\overline{P}_0\psi_n
+Q_\beta^2\psi_{n-1}+\ldots+Q_\beta^{n+1}\psi_0\bigr)\;,&&\nonumber\\
\qquad (n=0,1,\ldots)\;,&&
\label{e3.10}\\
(Q_\beta^0)^2\psi_n
=-Q_\beta^0\Bigl(
\bigl(-(\kappa+\frac{3}{2}(n-1))\widehat{Q}_\beta^1+
Q_\beta^1\bigr)\psi_{n-1}
+Q_\beta^2\psi_{n-2}+\ldots+Q_\beta^n\psi_0\Bigr)\;,&&\nonumber\\
\qquad (n=1,2,\ldots)&&
\label{e3.11}
\end{eqnarray}
(we used $P_0\widehat{Q}_\beta^1\overline{P}_0=0$). Here, and until the end of
this subsection, no summation over $\beta$ is understood. The equation
(\ref{e3.10}) at $n=0$ reads
\begin{equation} 
P_0Q_\beta^1\psi_0=\kappa P_0\widehat{Q}_\beta^1\psi_0
\;(=\kappa\widehat{Q}_\beta^1\psi_0)\;.
\label{e3.13}
\end{equation}
\par
We shall verify this by explicit computation later on. Since a similar issue
will show up in solving the equation (\ref{e3.10}) at $n>0$, let us also 
present a more general statement, whose proof is postponed to the next 
subsection.

\smallskip\noindent
{\bf Lemma} {\it Let $T_\beta$ be linear operators on the range of $P_0$, which
transform as real spinors of ${\rm Spin}(d)$ and commute with the antipode 
map. Then, for each invariant state we have
\begin{equation}
T_\beta\psi_0=\kappa\widehat{Q}_\beta^1\psi_0\;,
\label{e3.14}
\end{equation}
with $\kappa$ depending only on the associated representation (\ref{e3.8}).}

\medskip\noindent
We now assume having solved the equations (\ref{e3.10}, \ref{e3.11}) up to 
$n-1$ for ${\rm Spin}(d)$ invariant 
$\psi_1,\ldots\psi_{n-1}$ (which is true for $n-1=0$), and claim the
same is possible for $n$. Since $Q_\beta^0$ is
invertible on the range of $\overline{P}_0$, eq.~(\ref{e3.11})${}_n$
determines $\overline{P}_0\psi_n$ uniquely. The fact that
the solution so obtained is independent of $\beta$ and is ${\rm Spin}(d)$ 
invariant may deserve a comment, because the equivalence of the equations
$Q_\beta\psi=0$ and $(Q_\beta)^2\psi=0$, which holds on (\ref{e1.0}), does not
apply in the sense of formal power series (\ref{e2.4}). Consider the expansion
(\ref{e3.00}), i.e.,
\begin{equation*}  
Q_\beta = r^{1/2}\sum_{k=0}^\infty r^{-\frac{3}{2}k}[Q_\beta]_k\;, \qquad
[Q_\beta]_k= 
Q_\beta^k+\delta_{1k}\widehat{Q}_\beta^1r{\partial\over\partial r}\;,
\end{equation*}
as well as its formal square
\begin{equation*} 
(Q_\beta)^2 = r\sum_{k=0}^\infty r^{-\frac{3}{2}k}[(Q_\beta)^2]_k\;.
\end{equation*}
Notice that $(Q_\beta)^2$ is, by (\ref{eq2}), independent of $\beta$ and 
${\rm Spin}(d)$ invariant as an operator on ${\rm SU}(2)$ invariant power 
series. Similarly, let $[Q_\beta\psi]_k$ (given by
the l.h.s. of (\ref{e3.0})) and $[(Q_\beta)^2\psi]_k$ be the 
coefficients of the
corresponding series. By induction assumption we have $[Q_\beta\psi]_k=0$ for
$k=0,\ldots,\,n-1$. Since $Q_\beta(Q_\beta\psi)=(Q_\beta)^2\psi$,
we obtain
\begin{eqnarray*} 
[(Q_\beta)^2\psi]_n&=&\sum_{k=0}^n Q_\beta^k[Q_\beta\psi]_{n-k}
-(\kappa+\frac{3}{2}n-2)\widehat{Q}_\beta^1[Q_\beta\psi]_{n-1}
=Q_\beta^0[Q_\beta\psi]_n\;,\\{}
[(Q_\beta)^2\psi]_n&=&(Q_\beta^0)^2\psi_n+\widetilde{\psi}_{n-1}\;,
\end{eqnarray*} 
where $\widetilde{\psi}_{n-1}$ (determined by $\psi_0,\ldots\psi_{n-1}$) has
the desired properties. The equation (\ref{e3.11})${}_n$, i.e.,
$Q_\beta^0[Q_\beta\psi]_n=0$ is thus equivalent to 
$(Q_\beta^0)^2\psi_n=-\widetilde{\psi}_{n-1}$, which exhibits the claim.

On the other hand, invariance requires 
$P_0\psi_n$ to be a linear combination of invariant singlets. For the ansatz
$P_0\psi_n=\lambda_n\psi_0$,
eq. (\ref{e3.10})${}_n$ reads
\begin{equation} 
\frac{3}{2}n\lambda_n\widehat{Q}_\beta^1\psi_0=
-P_0\bigl(Q_\beta^1\overline{P}_0\psi_n
+Q_\beta^2\psi_{n-1}+\ldots+Q_\beta^{n+1}\psi_0\bigr)\;,
\nonumber
\end{equation}
because of (\ref{e3.13}). Again, by the lemma, this holds true for suitable 
$\lambda_n$. Indeed, this solution for $P_0\psi_n$ is the only one.

\subsection{Proof of the lemma}
\label{sec:pr5}

The vectors $T_\beta\psi_0,\, (\beta=1,\ldots, s_d)$ transform under 
${\rm Spin}(d)$ as real spinors,
although they might be linearly dependent. By reducing matters to the little
group as before, any representation of that sort is specified by the values 
$|F^\beta(E,\vec{e})\rangle$ of its states (see (\ref{e3.2}))
at one point $(E,\vec{e})$, which are required to satisfy
\begin{equation*} 
\widetilde{R}_{\beta\alpha}(R)|F^\alpha(E,\vec{e})\rangle=
\mathcal R(R)|F^\beta(E,\vec{e})\rangle
\end{equation*}
for $R$ with $RE=E$. Pretending the states $|F^\beta(E,\vec{e})\rangle$ to be
linearly independent, the branching
${\rm Spin}(d)\hookleftarrow{\rm Spin}(d-1)$ yields
\begin{eqnarray*}
16=8_{\rm s}\oplus 8_{\rm c}\quad(d=9)\;;&&\qquad 
4\oplus 4=(2_+\oplus 2_-)\oplus(2_+\oplus 2_-)\quad(d=5)\;;\\
2\oplus 2&=&(1_1\oplus 1_{-1})\oplus(1_1\oplus 1_{-1})\quad(d=3)\;.
\end{eqnarray*}
For $d=9,5$ each term on the r.h.s. occurs as often as in (\ref{e3.7}), and
$\psi_0$ can indeed be chosen so that the $s_d$ vectors 
$\widehat{Q}_\beta^1\psi_0$ 
are independent. Not so in the last case, where the vectors $T_\beta\psi_0$
just belong to $1_1\oplus 1_{-1}$. We
continue the discussion for different values of $d$ separately.
\par
$\bullet\,d=9$. Any linear transformation $K$ commuting with a 
${\rm Spin}(9)$ representation as above is thus of the form
$K=\kappa_{\rm s}\oplus \kappa_{\rm c}$.
If $K$ also commutes with the antipode map, then 
$\kappa_{\rm s}=\kappa_{\rm c}\equiv\kappa$. Applying this to the
representation $\widehat{Q}_\beta^1\psi_0$ and to the map 
$K:\,\widehat{Q}_\beta^1\psi_0\mapsto T_\beta\psi_0$ yields the
claim. 

$\bullet\,d=5$. Let us regroup 
$(2_+\oplus 2_-)\oplus(2_+\oplus 2_-)\cong 
(2_+\otimes\one_2)\oplus(2_-\otimes\one_2)$. Then any map $K$ commuting
with the representation is of the form
\begin{equation*}
K=(\one\otimes K_+)\oplus(\one\otimes K_-)\;,
\end{equation*}
where $K_-$ is conjugate to $K_+$ if $K$ commutes with the
antipode map. This allows for a four dimensional space of such maps $K$.
To proceed further we shall again assume that $E=(0,\ldots,0,1)$ and introduce
creation operators
\begin{equation*}
a_\alpha^*=\frac{1}{\sqrt{2}}[(\vec{\Theta}_\alpha\cdot\vec{e})+
\i(\vec{\Theta}_{\alpha+4}\cdot\vec{e})]\;,\qquad(\alpha=1,\ldots 4)
\end{equation*}
which then define a vacuum through $a_\alpha|0\rangle=0$. We next choose an
orthonormal basis $\{\psi_0^1,\ldots, \psi_0^4\}$ for the 4-dimensional 
subspace of singlets in the range of $P_0$ by specifying the values of the
corresponding fermionic parts (see (\ref{e3.2})) at 
$(E,\vec{e})$:
\begin{eqnarray*}
|F_0^4(E,\vec{e})\rangle&=&
\frac{1}{\sqrt{2}}(|0\rangle-a_1^*a_2^*a_3^*a_4^*|0\rangle)\;,\\
|F_0^i(E,\vec{e})\rangle&=&
\frac{1}{2\sqrt{2}}\widetilde{\Gamma}^i_{\alpha\beta}
a_\alpha^*a_\beta^*|0\rangle=
\frac{\i}{4}(\gamma^4\widetilde{\gamma}^i)_{\alpha\beta}
(\vec{\Theta}_\alpha\cdot\vec{e})(\vec{\Theta}_\beta\cdot\vec{e})
|F_0^4(E,\vec{e})\rangle\;,
\qquad(i=1,2,3)\;,\\
\end{eqnarray*}
where 
\begin{equation*}
\widetilde{\gamma}^i=\left( 
\begin{array}{cc}  0 & \i \widetilde{\Gamma}^i\\  
-\i \widetilde{\Gamma}^i& 0 \end{array} \right) 
=\sigma^{-1}\gamma^i\sigma\;,
\qquad
\sigma=\left( 
\begin{array}{cc} \Sigma & 0 \\ 0 & \Sigma\end{array} \right) 
\end{equation*}
with $\Sigma\in{\rm O}(4)$ and $\det\Sigma=-1$.
Note that $\psi_0^4$ is the singlet belonging to the 5--dimensional 
fermionic representation of ${\rm Spin}(5)$. One can verify that the four maps
\begin{equation*}
K^i:\,\widehat{Q}_\beta^1\psi_0^1\mapsto
\cases
\widehat{Q}_\beta^1\psi_0^i\;,&(i=1,2,3)\;,\\
\gamma^t_{\beta\alpha}E_t \widehat{Q}_\alpha^1\psi_0^4\;,&(i=4)\;,\\
\endcases
\end{equation*}
besides being of the kind just discussed, are linearly independent. Therefore 
any map $K$ of the above form is a linear combination thereof. In particular 
this applies,
for any $(\underline{x}, x_4)\in\mathbb R^{3+1}$, to the map
$K:\,\widehat{Q}_\beta^1\psi_0^1\mapsto 
x_iT_\beta\psi_0^i+x_4\gamma^t_{\beta\alpha}E_t T_\alpha\psi_0^4$,
hence
\begin{equation*}
x_iT_\beta\psi_0^i+x_4\gamma^t_{\beta\alpha}E_t T_\alpha\psi_0^4
=y_i\widehat{Q}_\beta^1\psi_0^i+
y_4\gamma^t_{\beta\alpha}E_t\widehat{Q}_\alpha^1\psi_0^4\;.
\end{equation*}
This defines a linear map 
$\kappa: (\underline{x},x_4)\mapsto(\underline{y},y_4)$
on $\mathbb R^{3+1}$. We claim that 
\begin{equation}
\kappa: (R\underline{x},x_4)\mapsto(R\underline{y},y_4)
\label{e3.13a}
\end{equation}
for 
$R\in {\rm SO}(3)$, which implies 
$\kappa={\rm diag}(\kappa_1=\kappa_2=\kappa_3, \kappa_4)$ and hence 
(\ref{e3.14}). Eq. (\ref{e3.13a}) can be proven using 
$R_{ij}\psi_0^i=\mathcal R\psi_0^j$ for $\mathcal R\in {\rm Spin}(8)$
projecting to $R\in{\rm Spin}(3)\subset{\rm Spin}(5)
\hookrightarrow{\rm SO}(8)$. This
in turn follows from (\ref{e1.2}) and from $\mathcal R\psi_0^4=\psi_0^4$.

$\bullet\,d=3$. Analogously to $d=9$.

\subsection{Determination of $\kappa$}
\label{sec:pr6}

Since $J_{AB}\psi_0=J_{st}\psi_0=0$ we may replace $Q_\beta^1$ by
\begin{equation} 
Q_\beta^1=\Theta_{\alpha A}\gamma^t_{\alpha\beta}
\bigl(-e_BE_tM_{BA}-e_AE_sM_{st}
-\frac{\i}{2}\,e_AE_ty_{sB}\frac{\partial}{\partial y_{sB}})+
\frac{1}{2}\vec{\Theta}_\alpha\cdot(\vec{y}_s\times\vec{y}_t)
\gamma^{st}_{\beta\alpha}\;.
\label{e4.1}
\end{equation}
We discuss the contributions to (\ref{e3.13}) of these four terms
separately. 
\par
i) With 
\begin{equation} 
e_BM_{BA}=-\frac{\i}{2}\bigl((\vec{\Theta}_\beta\cdot\vec{e})\Theta_{\beta A}-
\Theta_{\beta A}(\vec{\Theta}_\beta\cdot\vec{e})\bigr)
\nonumber
\end{equation}
we find
\begin{eqnarray*} 
\Theta_{\alpha A}e_BM_{BA}&=&\i\bigl(
(\vec{\Theta}_\alpha \cdot\vec{n}_+)(\vec{\Theta}_\beta\cdot\vec{n}_-)+
(\vec{\Theta}_\alpha \cdot\vec{n}_-)(\vec{\Theta}_\beta\cdot\vec{n}_+)
\bigr)(\vec{\Theta}_\beta\cdot\vec{e}\,)\;,\\
P_0\Theta_{\alpha A}e_BM_{BA}\psi_0&=&
\i(\vec{\Theta}_\alpha\cdot\vec{e}\,)\psi_0\;,
\end{eqnarray*} 
since only the term with $\beta=\alpha$ survives the projection $P_0$. Hence
\begin{equation} 
-P_0\Theta_{\alpha A}\gamma^t_{\alpha\beta}e_BE_tM_{BA}\psi_0
=\widehat{Q}_\beta^1\psi_0
\label{e4.3}
\end{equation}
contributes 1 to $\kappa$.
\par
ii) Similarly,
\begin{equation} 
-P_0(\vec{\Theta}_\alpha\cdot\vec{e}\,)\gamma^t_{\alpha\beta}E_sM_{st}\psi_0=
-(\vec{\Theta}_\alpha\cdot\vec{e}\,)\gamma^t_{\alpha\beta}E_s
M_{st}^{\parallel}\psi_0\;,
\nonumber
\end{equation}
where $M_{st}^{\parallel}$ is given in (\ref{e3.8a}). For the r.h.s. we then
claim
\begin{equation} 
-(\vec{\Theta}_\alpha\cdot\vec{e}\,)\gamma^t_{\alpha\beta}E_s
M_{st}^{\parallel}\psi_0=\kappa'\widehat{Q}_\beta^1\psi_0
\label{e4.3a}
\end{equation}
with 
\begin{equation} 
\kappa'=\cases 9\,,&\qquad(d=9)\;,\\
                    0,0,0,4\,,&\qquad(d=5)\;,\\
                    0,0\,,&\qquad(d=3)\;.
                     \endcases
\label{e4.3b}
\end{equation}
This is clear in the cases where the representation in (\ref{e3.8}) is 
already a singlet, i.e., when $\kappa'=0$. To prove the two 
remaining cases we first establish
\begin{equation} 
-(\vec{\Theta}_\alpha\cdot\vec{e}\,)\gamma^t_{\alpha\beta}E_s
M_{st}^{\parallel}\psi_0
=-\frac{\i}{2}\gamma^s_{\alpha\beta}E_s[\vec{\Theta}_\alpha\cdot\vec{e}\,,
M_{ut}^{\parallel}M_{ut}^{\parallel}]\psi_0
-\i\frac{d^2-d}{8}(\vec{\Theta}_\alpha\cdot\vec{e}\,)
\gamma^s_{\alpha\beta}E_s\psi_0\;,
\label{e4.4}
\end{equation}
or the equivalent equation obtained by multiplication from the right 
with $E_u\gamma^u$:
\begin{equation} 
-(\vec{\Theta}_\alpha\cdot\vec{e}\,)(\gamma^t\gamma^u)_{\alpha\beta}
E_uE_sM_{st}^{\parallel}\psi_0=
-\frac{\i}{2}[\vec{\Theta}_\beta\cdot\vec{e}\,,
M_{ut}^{\parallel}M_{ut}^{\parallel}]\psi_0
-\i\frac{d^2-d}{8}(\vec{\Theta}_\beta\cdot\vec{e}\,)\psi_0\;.
\label{e4.5}
\end{equation}
To this end
we note that, by the invariance of $\psi_0$, its
fermionic part $|F(E,\vec{e})\rangle$ at $E\in S^{d-1}$ is invariant under
rotations of ${\rm Spin}(d) $ leaving $E$ fixed:
$(\delta_{us}-E_uE_s)M_{sv}^{\parallel}(\delta_{vt}-E_vE_t)\psi_0=0$,
i.e., 
\begin{equation} 
(M_{st}^{\parallel}E_uE_s+M_{uv}^{\parallel}E_vE_t)\psi_0=
M_{ut}^{\parallel}\psi_0\;.
\label{e4.5a}
\end{equation} 
Using 
$\gamma^t\gamma^u=-\gamma^{ut}+\delta^{ut}\one$ and the observation just made
we rewrite the l.h.s. of (\ref{e4.5}) as 
\begin{eqnarray*}
-(\vec{\Theta}_\alpha\cdot\vec{e}\,)(\gamma^t\gamma^u)_{\alpha\beta}
E_uE_sM_{st}^{\parallel}\psi_0&=&
(\vec{\Theta}_\alpha\cdot\vec{e}\,)
\gamma^{ut}_{\alpha\beta}E_uE_sM_{st}^{\parallel}\psi_0\cr
&=&\frac{1}{2}(\vec{\Theta}_\alpha\cdot\vec{e}\,)\gamma^{ut}_{\alpha\beta}
(E_uE_sM_{st}^{\parallel}-E_tE_sM_{su}^{\parallel})\psi_0\cr
&=&\frac{1}{2}(\vec{\Theta}_\alpha\cdot\vec{e}\,)\gamma^{ut}_{\alpha\beta}
M_{ut}^{\parallel}\psi_0\;.
\end{eqnarray*}
The commutation relation 
\begin{equation*} 
\i[\vec{\Theta}_\alpha\cdot\vec{e}\, ,M_{ut}^{\parallel}]=
\frac{1}{2}\gamma^{ut}_{\alpha\beta}(\vec{\Theta}_\beta\cdot\vec{e}\,)
\end{equation*}
follows from (\ref{e1.2}) or by direct computation. It implies
\begin{eqnarray*}
\i[\vec{\Theta}_\alpha\cdot\vec{e}\, ,M_{ut}^{\parallel}M_{ut}^{\parallel}]&=&
\frac{1}{2}\gamma^{ut}_{\alpha\beta}\{
\vec{\Theta}_\beta\cdot\vec{e}\, ,M_{ut}^{\parallel}\}=
\gamma^{ut}_{\alpha\beta}(\vec{\Theta}_\beta\cdot\vec{e}\,)M_{ut}^{\parallel}
-\frac{1}{2}\gamma^{ut}_{\alpha\beta}
[\vec{\Theta}_\beta\cdot\vec{e}\,,M_{ut}^{\parallel}]\cr
&=&
\gamma^{ut}_{\alpha\beta}(\vec{\Theta}_\beta\cdot\vec{e}\,) M_{ut}^{\parallel}
-\i\frac{d^2-d}{4}\,\vec{\Theta}_\alpha\cdot\vec{e}\,\;.
\end{eqnarray*}
Solving for the first term on the r.h.s. proves (\ref{e4.5}) and hence 
(\ref{e4.4}). Let us now note that for $d=9$ the fermionic part of 
$\psi_0$, resp. of $(\vec{\Theta}_\alpha\cdot\vec{e}\,)\psi_0$ belongs to the 
$44$, resp. $128$ representation of ${\rm Spin}(9)$ (see (\ref{e3.6})). 
Eq. (\ref{e4.4}) then implies
\begin{equation*} 
-(\vec{\Theta}_\alpha\cdot\vec{e}\,)\gamma^t_{\alpha\beta}E_s
M_{st}^{\parallel}\psi_0=
(C(44)-C(128)+9)\widehat{Q}_\beta^1\psi_0=9\widehat{Q}_\beta^1\psi_0\;,
\end{equation*}
where we used the values \cite{b4} of the Casimir: $C(44)=C(128)=18$. In the case 
$d=5$ the
fermionic part of $\psi_0$, resp. of 
$(\vec{\Theta}_\alpha\cdot\vec{e}\,)\psi_0$ belongs to the representation
$5$, resp. $4\oplus 4$. We conclude that
\begin{equation*} 
-(\vec{\Theta}_\alpha\cdot\vec{e}\,)\gamma^t_{\alpha\beta}E_s
M_{st}^{\parallel}\psi_0=
(C(5)-C(4)+\frac{5}{2})\widehat{Q}_\beta^1\psi_0=4\widehat{Q}_\beta^1\psi_0\;,
\end{equation*}
given that  $C(5)=4,\,C(4)=5/2$.
\par
We remark that the proof of (\ref{e4.3b}) can be shortened by using the 
lemma, according to
which (\ref{e4.3a}) holds true for some $\kappa'$.
Thus, contracting with 
$\widehat{Q}_\beta^1\psi_0$ and summing over $\beta$, we find
\begin{eqnarray*}
-\kappa'(\psi_0,\widehat{Q}_\beta^1\widehat{Q}_\beta^1\psi_0)&=&
-\i(\psi_0,(\vec{\Theta}_\gamma\cdot\vec{e})\gamma^u_{\gamma\beta}E_u
(\vec{\Theta}_\alpha\cdot\vec{e})\gamma^t_{\alpha\beta}E_s
M_{st}^{\parallel}\psi_0)\\
&=&4(\psi_0,E_uM_{ut}^{\parallel}M_{st}^{\parallel}E_s\psi_0)\\
&=&2(\psi_0,M_{ut}^{\parallel}
(M_{st}^{\parallel}E_uE_s+M_{uv}^{\parallel}E_vE_t)\psi_0)
=2(\psi_0,M_{ut}^{\parallel}M_{ut}^{\parallel}\psi_0)\;.
\end{eqnarray*}
In the step before last we relabeled indices in half the expression; in the 
last one we used (\ref{e4.5a}).
Using $\widehat{Q}_\beta^1\widehat{Q}_\beta^1=-s_d/2$ we obtain 
$(s_d/2)\kappa'=2\cdot 2\cdot C$, i.e., $\kappa'=8C/s_d$,
where $C$ is the Casimir 
in the representation (\ref{e3.8}). The above values of $C(44)\,(d=9)$ and of 
$C(5)\,(d=5)$ yield again (\ref{e4.3b}). 
\par
iii) Using $\d\e^{-y^2/2}/\d y=
-y\e^{-y^2/2}$ we get
\begin{equation} 
\frac{1}{2}y_{sB}\frac{\partial}{\partial y_{sB}}\psi_0=
-\frac{1}{2}y_{sB}y_{sB}\psi_0=
-\frac{1}{2}\sum_{sB}(y_{sB}^2-\frac{1}{2})\psi_0-
\frac{1}{4}\cdot 2(d-1)\psi_0\;,
\label{e4.9}
\end{equation}
where the sum, consisting of second Hermite functions, is annihilated by
$P_0$.
\par
iv) The last term in (\ref{e4.1}), when acting on $\psi_0$, is similarly 
annihilated by $P_0$.

\smallskip
Collecting terms (\ref{e4.3}, \ref{e4.3b}, \ref{e4.9}) we find
\begin{equation*} 
\kappa=1+\kappa'-\frac{1}{2}(d-1)=\cases 6\,,&\qquad(d=9)\;,\\
                              -1,-1,-1,3\,,&\qquad(d=5)\;,\\
                             0, 0\,,&\qquad(d=3)\;.
                \endcases
\end{equation*}

\setcounter{section}{0}
\section*{Appendix 1}

To prove (\ref{e3.000}) we shall compute the partial derivatives in 
\begin{equation}
\frac{\partial}{\partial q_{tA}}=
\frac{\partial r}{\partial q_{tA}}\frac{\partial}{\partial r}+
\frac{\partial e_B}{\partial q_{tA}}\frac{\partial}{\partial e_B}+
\frac{\partial E_s}{\partial q_{tA}}\frac{\partial}{\partial E_s}+
\frac{\partial y_{sB}}{\partial q_{tA}}\frac{\partial}{\partial y_{sB}}\;.
\label{eq:a0}
\end{equation}
We regard $r,\,\vec{e},\,E,\,y$ as functions of $q$ defined by 
$\vec{e}{\,}^2=\sum_s E_s^2 = 1$ and (\ref{e2.00}, \ref{e2.1})
and solve for their differentials by taking different contractions of
\begin{equation*}
\d q_{tA}=(e_AE_t-\frac{1}{2}r^{-3/2}y_{tA})\d r+
rE_t\d e_A+re_A\d E_t+r^{-1/2}\d y_{tA}\;.
\end{equation*}
Using that
\begin{eqnarray*}
e_A\d y_{tA}+y_{tA}\d e_A=0\;,&\qquad E_t\d y_{tA}+y_{tA}\d E_t=0\;,&\\
e_A\d e_A=0\;,&\qquad E_t\d E_t=0\;,&
\end{eqnarray*}
the contractions are:
\begin{eqnarray}
e_AE_t\d q_{tA}&\!=\!&\d r\;,\nonumber\\
(\delta_{BA}-e_Be_A)E_t\d q_{tA}&\!=\!&r\d e_B-r^{-1/2}y_{tA}\d E_t\;,
\label{eq:a1}\\
e_A(\delta_{st}-E_sE_t)\d q_{tA}&\!=\!&r\d E_s-r^{-1/2}y_{sA}\d e_A\;,
\label{eq:a2}\\
(\delta_{BA}-e_Be_A)(\delta_{st}-E_sE_t)\d q_{tA}&\!=\!&
-\frac{1}{2}r^{-3/2}y_{sB}\d r+
r^{-1/2}(\d y_{sB}+e_By_{sA}\d e_A+E_sy_{tB}\d E_t)\;.
\nonumber
\end{eqnarray}
We solve (\ref{eq:a1}, \ref{eq:a2}) for $\d e_B,\, \d E_s$:
\begin{eqnarray*}
\d r&=&e_AE_t\d q_{tA}\;,\\
\d e_B&=&(m^{-1})_{BC}(r^{-1}(\delta_{CA}-e_Ce_A)E_t+r^{-5/2}y_{tC}e_A)
\d q_{tA}\\
&=&(r^{-1}(\delta_{BA}-e_Be_A)E_t+{\rm O}(r^{-5/2}))\d q_{tA}\;,\\
\d E_s&=&(M^{-1})_{su}(r^{-1}(\delta_{ut}-E_uE_t)e_A+r^{-5/2}y_{sA}E_t)
\d q_{tA}\\
&=&(r^{-1}(\delta_{st}-E_sE_t)e_A+{\rm O}(r^{-5/2}))\d q_{tA}\;,\\
\d y_{sB}&=&[r^{1/2}(\delta_{BA}-e_Be_A)(\delta_{st}-E_sE_t)+
\frac{1}{2}r^{-1}e_AE_ty_{sB}]\d q_{tA}
-e_By_{sA}\d e_A-E_sy_{tB}\d E_t\;,
\end{eqnarray*}
where $m,\, M$ are the matrices
\begin{equation*}  
m_{AB}=\delta_{AB}-r^{-3}y_{tA}y_{tB}\;,\qquad
M_{st}=\delta_{st}-r^{-3}y_{sA}y_{tA}\;.
\end{equation*}  
We can now read off the partial derivatives appearing in (\ref{eq:a0}) and
obtain
\begin{eqnarray}
\frac{\partial}{\partial q_{tA}}&=&
r^{1/2}(\delta_{st}-E_sE_t)(\delta_{AB}-e_Ae_B)
\frac{\partial}{\partial y_{sB}}
+r^{-1}[e_AE_t(r\frac{\partial}{\partial r}+
\frac{1}{2}y_{sB}\frac{\partial}{\partial y_{sB}})]
\nonumber\\
&&+r^{-1}(\delta_{AC}-e_Ae_C)E_t
(\delta_{CB}\frac{\partial}{\partial e_B}-e_By_{sC}\frac{\partial}{\partial
y_{sB}}) \nonumber\\
&&+r^{-1}(\delta_{ut}-E_uE_t)e_A
(\delta_{us}\frac{\partial}{\partial E_s}-E_sy_{uB}\frac{\partial}{\partial
y_{sB}})
+{\rm O}(r^{-5/2})\;,\label{eq:a4}
\end{eqnarray}
with the remainder not containing derivatives w.r.t. $r$. Finally, we insert
this expression into 
\begin{eqnarray*}
\i L_{BA}&=&
q_{sB}\frac{\partial}{\partial q_{sA}}-q_{sA}\frac{\partial}{\partial q_{sB}}\\
&=&[(\delta_{AC}-e_Ae_C)y_{sB}-(\delta_{BC}-e_Be_C)y_{sA}]
\frac{\partial}{\partial y_{sC}}\\
&&+e_B(\delta_{AC}\frac{\partial}{\partial e_C}-
e_Cy_{sA}\frac{\partial}{\partial y_{sC}})
-e_A(\delta_{BC}\frac{\partial}{\partial e_C}-
e_Cy_{sB}\frac{\partial}{\partial y_{sC}})\;,
\end{eqnarray*}
(with no higher order corrections, as $L_{AB}$ is of exact order ${\rm O}(r^0)$) and 
then into
\begin{equation*}
\i r^{-1} e_BE_tL_{BA}=r^{-1}(\delta_{AC}-e_Ae_C)E_t
(\delta_{CB}\frac{\partial}{\partial e_B}-e_By_{sC}\frac{\partial}{\partial
y_{sB}})\;.
\end{equation*}
Similarly, we have
\begin{equation*}
\i r^{-1} e_AE_sL_{st}=r^{-1}(\delta_{ut}-E_uE_t)e_A
(\delta_{us}\frac{\partial}{\partial E_s}-E_sy_{uB}\frac{\partial}{\partial
y_{sB}})\;.
\end{equation*}
Together with (\ref{eq:a4}), this proves (\ref{e3.000}).

\section*{Appendix 2}

Consider
\begin{equation}  \label{xy-Hamiltonian}
H = ( -{\partial_x}^2 -{\partial_y}^2 + x^2y^2 ) \one + 
\left( 
\begin{array}{cc}
x & -y \\
-y & -x
\end{array}
\right)\;,
\end{equation}
which is the square of
\begin{equation*}      
Q = \i 
\left( 
\begin{array}{cc}
\partial_x  & \partial_y + xy \\
\partial_y - xy  & - \partial_x
\end{array}
\right) .
\end{equation*}
Just as in (\ref{e1.4}), the bosonic potential $V$ ($=x^2y^2$) is
non-negative, 
but vanishing in regions of the configuration space that extend to infinity
(causing the classical partition function to diverge). Quantum--mechanically,
just as in (\ref{e1.4}), the bosonic system is stabilized by the zero point
energy of fluctuations transverse to the flat directions; the fermionic matrix
part in (\ref{xy-Hamiltonian}) exactly cancels this effect, causing the
spectrum to cover the whole positive real axis \cite{c4b}. As simple as it is,
it has remained an open question (for now more than 10 years) whether
(\ref{xy-Hamiltonian}) admits a normalizable zero energy solution, or
not. The argument, derived in a few lines below, gives `no' as an answer
and provides the simplest illustration of our method: as $x \rightarrow
+ \infty, \ Q \Psi = 0$ has two approximate solutions,
\begin{equation} \label{xy-solutions}
\Psi_{+} = e^{-\frac{xy^2}{2}} \left( \begin{array}{c} 0 \\ 1 \end{array}
\right)  
\quad \mathrm{and} \quad
\Psi_{-} = e^{+\frac{xy^2}{2}} \left( \begin{array}{c} 1 \\ 0 \end{array}
\right)\;,
\end{equation}
the first of which should be chosen for $\Psi_{0}$ in the
asymptotic expansions
\begin{equation}     \label{xy-expansion}
\Psi = x^{-\kappa} ( \Psi_{0} + \Psi_{1} + ...  )\; .
\end{equation}
In this simple example, the sum $Q = \sum_{n=0}^{\infty} Q^{(n)}$ terminates
after the first two terms, and
\begin{equation*}
0 \stackrel{\mathrm{!}}{=} Q \Psi = 
\left( \left( 
\begin{array}{cc}
0 & \partial_y + xy  \\
\partial_y -xy  & 0
\end{array}
\right)
+
\left( 
\begin{array}{cc}
\partial_x  & 0 \\
0  & - \partial_x
\end{array}
\right) \right)
\left( x^{- \kappa} ( \Psi_0 + \Psi_1 + ...  ) \right)\;,
\end{equation*}
yields (as already anticipated, cp. (\ref{xy-solutions}))
\begin{equation*}       
\left( 
\begin{array}{cc}
0 & \partial_y + xy \\
\partial_y -xy & 0
\end{array}
\right)
\Psi_0 = 0
\end{equation*}
and 
\begin{equation}  \label{xy-eqn2}
\left( 
\begin{array}{cc}
0 & \partial_y + xy \\
\partial_y -xy & 0
\end{array}
\right) \Psi_n +
x^{\kappa} 
\left( 
\begin{array}{cc}
\partial_x & 0 \\
0  & - \partial_x
\end{array}
\right) x^{-\kappa} \Psi_{n-1} = 0\;, \qquad n = 1, 2, ...  \; .
\end{equation}
Multiplying (\ref{xy-eqn2}) by $\Psi_{0}^{\dagger}$ and integrating over $y$
one 
sees that
\begin{equation*}
\int_{- \infty}^{+ \infty} e^{-\frac{xy^2}{2}} x^{\kappa}( 0 , -
\partial_x) x^{-\kappa} \Psi_{n-1} dy
\end{equation*}
has to vanish, implying in particular 
\begin{eqnarray*} 
0 & = & \int_{- \infty}^{+ \infty} \left( \frac{y^2}{2} + 
\frac{\kappa}{x} \right)
e^{- x y^2} 
dy\;,
\\  \nonumber
\kappa & = & - \frac{1}{4} \; ,
\end{eqnarray*}
which proves that (\ref{xy-Hamiltonian}) does not admit any 
square--integrable solution of the form (\ref{xy-expansion}). A different
approach has recently been undertaken by Avramidi \cite{avr}. Finally note
that, calculating the $\Psi_{n > 0}$ 
from 
(\ref{xy-eqn2}), yields the asymptotic expansion, $x \rightarrow +
\infty $, 
\begin{equation*}
\Psi(x,y) = 
x^{\frac{1}{4}} e^{-\frac{xy^2}{2}} \sum_{n=0}^{\infty} x^{-
\frac{3n}{2}}
\left( 
\begin{array}{c}
\frac{y}{4x} f_n(xy^2) \\
 g_n(xy^2)
\end{array}
\right)\;,
\end{equation*}
where $f_0 = 1 = g_0,\, f_1=0=g_1$, and the $f_n  (s), \,g_{n}(s)$ are the
(unique) polynomial solutions
\begin{equation*}
f_n (s) = \sum_{i = 0}^n f_{n,i} s^i\; , \qquad
g_n(s) = \sum_{ i = 0}^n g_{n,i} s^i
\end{equation*}
of
\begin{eqnarray*}
2s f'_n + ( 1 - 2s ) f_n & = & \left( 1 - 2s - 6n \right) g_n + 4s g'_n\;, 
\\ \nonumber
8 g'_{n+2} & = & \left( \frac{3}{4} + \frac{s}{2} + \frac{3n}{2} \right) 
f_n - s f'_n\; .
\end{eqnarray*}

\medskip
\noindent {\bf Acknowledgments.\/} We thank A. Alekseev, I. Avramidi, V. Bach, 
F. Finster, H. Nicolai, C. Schweigert, R. Suter, P. Yi for useful discussions.
We also
thank the following institutions for support: the Albert Einstein Institute,
the Fields Institute,
the Erwin Schr\"odinger Institute, the Institute for Theoretical
Physics of ETH, the Mathematics Department of Harvard University, the Deutsche
Forschungsgemeinschaft.

\end{document}